\documentstyle[12pt,amssymb]{article}

\newcommand{\be}{\begin{equation}}
\newcommand{\bel}[1]{\begin{equation}\label{#1}}
\newcommand{\ee}{\end{equation}}
\newcommand{\bea}{\begin{eqnarray}}
\newcommand{\ba}{\begin{array}}
\newcommand{\eea}{\end{eqnarray}}
\newcommand{\ea}{\end{array}}
\newcommand{\hfour}{\hspace*{4mm}}
\newcommand{\bra}[1]{\mbox{$\langle \, {#1}\, |$}}
\newcommand{\ket}[1]{\mbox{$| \, {#1}\, \rangle$}}
\newcommand{\exval}[1]{\mbox{$\langle \, {#1}\, \rangle$}}

\begin{document}
\begin{titlepage}             %
\thispagestyle{empty}         %
\begin{center}                %
{\large                       
{\bf
Diffusion-limited annihilation in inhomogeneous environments\\
}
}\vspace{3cm}                 %
{\large {\sc                  
Gunter M. Sch\"utz
}}\\[8mm]                     %
{
Institut f\"ur Festk\"orperforschung, Forschungszentrum J\"ulich,\\
52425 J\"ulich, Germany
}
\vspace{2cm}\\                %
\begin{minipage}{10cm}{\small\sl\rm 
We study diffusion-limited (on-site) pair annihilation $A+A\to 0$ and (on-site)
fusion $A+A\to A$ which we show to be equivalent for arbitrary 
{\em space-dependent}
diffusion and reaction rates. For one-dimensional 
lattices with nearest neighbour hopping we find that in the limit of 
infinite reaction rate the time-dependent $n$-point 
density correlations for {\em many-particle} initial states are
determined by the correlation functions of a dual diffusion-limited
annihilation process with at most $2n$ 
particles initially. Furthermore, by reformulating general properties of
annihilating random walks in one dimension in terms of fermionic anticommutation
relations we derive an exact representation for these correlation functions in 
terms of conditional probabilities for a {\em single} 
particle performing a random walk with dual hopping rates. 
This allows for the exact and explicit calculation of a wide range of universal 
and non-universal types of behaviour for the decay of the density and density 
correlations.
}
\end{minipage}
\end{center}
\end{titlepage}           %
\baselineskip 0.3in       %

\section{Introduction}

In this paper we investigate a class of models describing diffusion-limited
annihilation of identical particles. This is a process where particles
of a single species diffuse (rather than move ballistically) and undergo
an annihilation reaction when two particles meet. The outcome of such a
reaction may be either a single particle, or both particles disappear.
This process can describe both chemical reactions where the particles change
their state into an inert reaction product which takes no part in the
subsequent dynamics of the system, or physical reactions where the particles
actually annihilate under the emission of radiation. On an abstract level,
both kinds of reactions are identical. A known one-dimensional process with 
``physical'' annihilation is e.g. laser-induced exciton dynamics on polymers 
\cite{Priv}. Clearly, one is not interested in the equilibrium behaviour of 
such a model, since in the absence of continued particle production all 
particles eventually disappear. Instead one would like to understand 
non-equilibrium properties such as the time-dependence of the particle 
concentration for a given initial condition and the degree of universality
obtained from specific choices of models.

Diffusion-limited pair annihilation of particles of a single species 
is a well-studied system (for a recent review see \cite{Priv}), but there are 
still interesting open questions particularly in the presence of spatially
inhomogeneous hopping and reaction rates 
which have so far not been addressed. In homogeneous, translationally
invariant environments the particle density decays with a power law which
depends on the dimensionality of the system. Both theoretically and
experimentally one finds $\rho(t) \sim 1/\sqrt{Dt}$ in one dimension. 
Interestingly, this result is at variance with the 
dimensionality-{\em in}dependent mean-field behaviour $\rho(t) \sim 1/(Dt)$ 
which is correct only in three (and higher) dimensions.
The amplitude of this decay is universal in the sense that it depends neither on
the initial density for random initial conditions nor on the reaction 
rate \cite{Lee94}.
However, if particles are moving in an arbitrary, non-translationally invariant 
energy landscape it is not obvious how this will change the decay of the local 
or overall particle concentration. In our study which is a continuation
of previous work \cite{Schu97} we first discuss universality of a general
class of models, but then
shall pay special attention to one-dimensional systems with nearest neighbour
interaction. The physical motivation behind the study of one-dimensional
system is not only their experimental relevance for polymer physics, but
also their theoretical importance in the understanding of the role of
fluctuations in low-dimensional systems. In both one- and two-dimensional
systems diffusive mixing is inefficient and leads to the building up
of large-scale correlations. Thus the classical mean-field rate equations for 
the study of these systems tend to fail and require a more sophisticated
treatment. 

To this end we investigate by exact methods a family of lattice models
with space-dependent hopping and pair annihilation
rates, as one would have e.g. in disordered systems or in the presence
of a space-dependent external forces affecting the diffusive motion of the
particles. These particles have no attractive or repulsive interaction
between themselves, they hop with fixed rates $d_{xy}$ from lattice site $x$
to site $y$. When two particles meet on site $x$ they both annihilate
with a rate $\lambda_x$. We address two different problems: (1) The first is 
to identify other systems which are in the same universality class in 
any dimension. In fact, our approach of using similarity transformations 
extends to systems defined on arbritrary lattices. This idea is not new
for systems with exclusion dynamics \cite{Stin93} - \cite{Simo95}. 
Here we extend the application of similarity transformations to models 
without site-exclusion. This generalization is useful for a renormalization 
group treatment of the problem \cite{Peli85,Lee94,Card96} as it
yields exact relations for expectation values for one model in terms 
of expectation values for the related system without any further RG analysis for 
each of these different models. (2) Since the RG treatment shows that
the fixed point of the homogeneous system with nearest neighbour hopping is the 
limit of infinite reaction rate, and since mean-field is particularly poor
in one dimension, the second problem we address is a more 
detailed investigation of the one-dimensional case in the limit of infinite
reaction rate but with (arbitrary) space-dependent nearest neighbour hopping 
rates. 
Here we use not only similarity transformations but also free fermion techniques
which, as shown below, are a convenient framework to formulate general 
properties of annihilating random walks in one dimension.
From our treatment we find that the time-dependent 
density correlations for an arbitrary {\em many-particle} initial state have an 
exact representation in terms of conditional probabilities for a {\em single} 
particle performing a random walk with dual hopping rates. 
In the case of constant hopping rates the predictions of this
model are in excellent agreement with experimental data on exciton dynamics
on very long ordered polymer chains. Thus we expect that our model 
gives an equally good description of the behaviour of inhomogeneous 
systems.

The paper is organized as follows. In Sec. 2 we briefly review the description
of reaction-diffusion processes in terms of a master equation which is written
in a quantum Hamiltonian formalism. In Sec. 3 equivalences between models
without site exclusion are studied. This is extended in Sec. 4 to the limit
of infinite reaction rate and nearest neighbour hopping and the main results
of the paper are derived. 
In Sec. 5 some specific results for random initial conditions are derived
and in Sec. 6 the main results are summarized and
discussed.

\section{Quantum Hamiltonian formalism for stochastic reaction-diffusion 
systems}
\setcounter{equation}{0}

The reaction-diffusion processes considered in this work are
systems of {\em classical} interacting particles which hop stochastically
on a lattice and undergo some reaction when they meet on the same lattice site. 
For definiteness we shall consider only systems of one species
of particles even though most of what is discussed in this and the following
Section can be generalized to
systems of many species. In symbolic presentation particles are denoted by $A$, 
a vacant site by $\emptyset$. The expression $kA$ denotes the presence of 
$k$ particles are in the same site.

We define the process in terms of a master equation for the probability
$P(\eta;t)$ of finding, at time $t$, any configuration $\eta$
of particles on a lattice $S$ of ${\cal L}$ sites. Here 
$\eta = \{\eta(1), \eta(2), \dots , \eta({\cal L})\}$ where $\eta(x)$ are the 
integer-valued particle occupation numbers at site $x$. The lattice is at this 
stage arbitrary.
Later, when discussing one-dimensional systems we consider only linear chains
with ${\cal L}=L$ sites and periodic boundary conditions. For clarity
we shall use in this case site labels $k,l,m$ instead of $x,y,z$.

A general master equation reads
\bel{2-0}
\frac{d}{d t} P(\eta';t) = \sum_{\eta\neq \eta'} 
\left\{ w(\eta',\eta) P(\eta,t) - w(\eta,\eta') P(\eta';t)\right\} 
\ee
The positive contribution to the (infinitesimal) change in probability
is the sum of probabilities that the system has been in a state $\eta$ times
the rate  $w(\eta',\eta)$ of flipping to $\eta'$.
The rates satisfy $0 \leq w(\eta',\eta) < \infty$ and define the exponentially 
distributed life time $\tau^{-1}(\eta') \equiv \sum_{\eta\neq \eta'} 
w(\eta,\eta')$ of a state $\eta'$. This is the total rate for the state $\eta'$ 
to flip into any other configuration and hence represents the loss in 
probability for the state $\eta'$ in Eq. (\ref{2-0}). We shall now express the 
time evolution
given by the master equation in terms of a quantum Hamiltonian $H$.
This is  discussed in detail in \cite{sch} and also in earlier work 
\cite{Kada68,Doi76,Gras80}
and we shall repeat only the essential elements of the mapping.
The advantage of this approach is that there are standard methods of dealing
with the resulting time evolution operator $H$. The applicability of these 
techniques, in the case
at hand primarily similarity transformations and the renormalization group 
technique for systems without
site exclusion and free fermion techniques for models with site exclusion, 
does not arise naturally
if the master equation is written down in the standard form (\ref{2-0}).

\subsection{Definitions}

The idea is to represent each of the possible particle configurations $\eta$
in the set $X={\mathbb{N}}^{\cal L}$ by a
vector $\ket{\eta}$ which together with the transposed vectors 
$\bra{\eta}$ form an {\em orthonormal} basis 
of a vector space $X=({\mathbb{C}}^\infty)^{\otimes L}$. 
The basis is chosen such that for a single site the state $\ket{n}$ with 
$\eta(x)=n$ particles is represented by the unit column vector with a 1 at the 
$n^{th}$ position and zero elsewhere. Then the state of the lattice as a
whole is represented by the tensor product 
$\ket{\eta}=\ket{\eta(1)}\otimes \dots \otimes \ket{\eta({\cal L})}$.
Now one can conveniently represent the probability distribution by a state 
vector
\bel{2-1}
| \, P(t)\, \rangle = \sum_{\eta \in X} P(\eta;t) \ket{\eta}.
\ee
Using $P(\eta;t) = \langle \, \eta \, | \, P(t) \, \rangle$ one writes 
the master equation (\ref{2-0}) in the form
\bel{2-3}
\frac{d}{dt} P(\eta;t) = - \langle \, \eta \, |
H | \, P(t) \, \rangle 
\ee
where the off-diagonal matrix elements of $H$ are the (negative) transition 
rates and the diagonal entries are the life times of the states.

Therefore the time evolution of this probability vector is given
in terms of a linear 'Hamilton' operator $H$ acting on $X$
\bel{2-2}
\frac{d}{dt} | \, P(t)\, \rangle =
- H | \, P(t)\, \rangle .
\ee
A state at time $t=t_0 + \tau$ is given in terms of an initial
state at time $t_0$ by
\bel{2-2a}
| \, P(t_0+\tau) \, \rangle = \mbox{e}^{-H\tau } 
| \, P(t_0) \, \rangle .
\ee
Note that $\bra{s} \, P(t) \, \rangle = \sum_{\eta \in X} P(\eta;t) = 1\; 
\forall \, t$
where
\bel{2-4}
\langle \, s \,| = \sum_{\eta \in X} \langle \, \eta \, |
\ee
This relation expresses conservation of probability and implies 
\bel{2-5}
\bra{s} H = 0
\ee
for any stochastic process. We shall call a matrix which satisfies $(\ref{2-5})$
and $H_{\eta,\eta'} \leq 0$ for all off-diagonal elements a 
{\em stochastic Hamiltonian}.

This definition is manifestly basis dependent. If a stochastic Hamiltonian $H$
is related to some other Hamiltonian $\tilde{H}$ by a similarity transformation,
\bel{2-6a}
H = B \tilde{H} B^{-1}
\ee
we shall call the systems {\em equivalent}, irrespective of whether $\tilde{H}$ 
is stochastic or not. If $\tilde{H}$ is stochastic, then the two processes
are called equivalent. An equally useful relation is
\bel{2-6b}
H = B \tilde{H}^T B^{-1}
\ee
relating a stochastic process to the transpose of a matrix $\tilde{H}$.
If both $H$ and $\tilde{H}$ are stochastic, we shall say that these two
processes are {\em enantiodromic} with respect to each other. Any family of
transformations $B$ generates a family of related processes (equivalent
or enantiodromic). Any one member of this family is called a 
{\em representative}. 

Expectation values $\exval{Q(t)}= \bra{s} Q \ket{P(t)}$ are 
calculated as matrix elements 
of suitably chosen diagonal operators $Q$.
A complete set of observables are the occupation numbers $\eta(x)$.
Defining projection {\em operators} on states with $n$ particle on site $x$ of
the lattice as
\bel{2-7}
Q_x(n) = \ket{n}_x\bra{n}_x
\ee
one finds that the density $\rho_x(t)$ of particles at 
site $x$ is given by the expectation value $\exval{n_x}$ of the operator $n_x = 
\sum_{n=0}^\infty n Q_x(n)$. This is the diagonal matrix $n$ 
acting non-trivially only on site $x$ with matrix elements 
$(n)_{kl} = k \delta_{k,l}$. Density orrelation functions 
$\langle \, n_{x_1} \cdots n_{x_j} \, \rangle$ are computed analogously. 

For later convenience we also introduce the operators $a^\pm_x$. The 
single-site matrix $a^+$ has elements $(a^+)_{m,n} =  \delta_{m,n+1}$. Hence
\bel{2-8}
a^+ \ket{n} = \ket{n+1} \; , \; \bra{n} a^+ =  \bra{n-1}.
\ee
creates a particle with unit rate when acting to the right. The single-site 
matrix $a^-$ has elements $(a^-_x)_{m,n} = n \delta_{m,n-1}$. Thus
\bel{2-9}
a^- \ket{n} = n \ket{n-1} \; , \; \bra{n} a^- = (n+1) \bra{n+1}.
\ee
annihilates a particle, but with a rate proportional to the occupation number 
prior to the annihilation. This corresponds to annihilating any one of the
existing identical particles with unit rate. It is understood in these equations 
that 
$a^- \ket{0}=\bra{0} a^+ =0$. Note that the 
matrices $a^{\pm}$ satisfy a harmonic oscillator algebra $a^- a^+ - a^+a^-=1$.
The product $n = a^+a^-$ is the number operator. Matrices acting on
different sites commute.

The state $\bra{s}$ is the coherent state $\bra{s}= \bra{0} \exp{(A^-)}$ where
$A^\pm = \sum_x a^\pm_x$ and $\bra{0}\equiv (\bra{0})^{\otimes {\cal L}}$ is the 
row vector representing the completely empty lattice. Using the harmonic 
oscillator algebra and the factorization of $\exp{(A^-)}$ one finds the
following relations
\bel{2-10}
\bra{s} a^+_x = \bra{s} \; , \; \bra{s} a^-_x = \bra{s} n_x
\ee
which are useful for the construction of the Hamiltonian $H$ and for the
derivation of the equations of motion for correlation functions.

In the infinite reaction limit discussed below, the system we investigate 
reduces to a
two-state model where each lattice site is either empty or occupied
and the state space reduces to $X = ({\mathbb{C}}^2)^{\otimes {\cal L}}$.
For this situation is more convenient to use Pauli matrices $\sigma^\alpha_k$
acting on site $k$
for the description
of the elementary events of the stochastic time evolution. 
A complete set of observables are now the occupation numbers $n_k=0,1$.
Projectors on states with a particle on site $k$ of
the chain are then defined by the diagonal matrix
\bel{2-11}
n_k = \frac{1}{2} \left( 1 - \sigma^z_k \right).
\ee
One may equally well use spin language. In this interpretation $n_k$
projects on a spin down at site $k$ and $v_k \equiv 1-n_k$ projects on
vacancies or spin up respectively.

Changes in the occupation numbers are effected by
$s^{\pm}_k = (\sigma^x_k \pm i \sigma^y_k)/2$. 
In our convention $s^-_k$ creates a particle at site $k$ when acting to the 
right, while $s^+_k$ annihilates a particle at site $k$.\footnote{This somewhat
counterintuitive notation results from long standing convention which we do not
wish to change.} Note that
\bel{2-12} 
\langle \, s \,| s^+_k = \langle \, s \,| n_k
\hfour \mbox{and}  \hfour
\langle \, s \,| s^-_k = \langle \, s \,| (1-n_k )   .
\ee
Introducing the ladder operators $S^{\pm} = \sum_{k=1}^L s^{\pm}_k$
one may write
\bel{2-13}
\langle \, s \,| = \langle \, 0 \,| \, \mbox{e}^{S^+}   .
\ee
Using the commutation relations for the Pauli matrices then yields (\ref{2-12}).

Some particular initial states and initial distributions used below:
An initial state with $N$ particles place on sites $x_1,\dots,x_N$ is denoted
by $\ket{x_1,\dots,x_N}$. An uncorrelated initial state with a site-independent
Poisson distribution is denoted by $\ket{\rho}$ where $\rho$ stands 
collectively for all average site-densities $\rho_x$. The vector
representing this inhomogeneous product measure is the factorized state
$\ket{\rho} = \prod_x \ket{P_0(x)}$ where $\ket{P_0(x)}=\exp{[\rho_x(a^+_x -1)]}
\ket{0}$. For two-state models
the same notation is used for an uncorrelated initial state with bimodal
distribution. In this case, the probability of finding site $x$ occupied is 
$\rho_x$ and the probability of finding it vacant is $1-\rho_x$. The completely 
empty lattice is always represented by the vector $\ket{0}$.

\subsection{Construction of the quantum Hamiltonian}

Now we are in a position to derive $H$. The process we consider as
a representative is defined by following elementary rules: 
\begin{itemize}
\item Particles move on a lattice from sites
$x \to y$ with a fixed rate $p(x,y)$ which is not necessarily symmetric. 
\item Any two particles on a site can annihilate with rate
$\tilde{\lambda}(x)$.
\end{itemize}
The idea behind these rules is the description of particles which have no
physical interaction (repulsive or attractive), but only a chemical interaction
which allows the formation of an inert product from two single particles
which then has no influence on the subsequent dynamics. In this model
no assumptions are made about the physical or chemical origin of the space 
dependence of these rates.

The precise nature of the dynamics is defined by a master equation (\ref{2-0}).
With matrices defined above the stochastic quantum Hamiltonian for the
process has the form $H^I = H^I_D + H^I_R$ where $H^I_D$ describes hopping
$H^I_R=\sum_x h^I_x$ describes annihilation. It is given by
\be
\label{2-14}
H^I = - \sum_{x\neq y \in S} p(x,y) (a^-_x a^+_y - n_x) 
 - \sum_{x \in S} \tilde{\lambda}(x) \left( (a^-_x)^2 - n_x(n_x-1) \right).
\ee
The first sum runs over all pairs $(x,y)$ of sites of the lattice 
$S$ and describes hopping of non-interacting particles from $x$ to $y$ whereas 
the second sum runs over all sites of the lattice and gives the annihilation
events.
According to the general rules the off-diagonal part represents the possible
moves the system can perform in an infinitesimal time step. The diagonal part
ensures that probability is conserved. Using (\ref{2-10}) and the harmonic
oscillator algebra it is easy to check that (\ref{2-5}) holds for
each elementary process and hence for $H$.

Consider now the infinite annihilation limit $\tilde{\lambda}(x) = 
 \lambda(x)/\lambda_0$ with $\lambda_0 \to 0$. In practical terms this means 
that the time scale $\lambda_0$ on which the annihilation takes place is much 
shorter than the time scales $p(x,y)$ set by the hopping events. In this limit 
two things happen: Firstly,
in the initial state any site occupancy by an even number of particles is 
immediately reduced to a vacancy and any site with an odd number of particles 
is left with just a single particle. Secondly, if by a hopping
event a double occupancy is created, this results immediately in a pair of empty
sites. Thus an attempted hopping event of the form  
$A_x A_y \to \emptyset_x 2 A_y$ with rate $p(x,y)$ is equivalent to the pair 
annihilation process $A_x A_y \to \emptyset_x \emptyset_y$ with rate $p(x,y)$. 
In short, the system becomes equivalent to an
exclusion process with particle hopping and pair annihilation with annihilation
rates that are equal to the sum of the hopping rates $p(x,y)+p(y,x)$. 
The Hamiltonian for this process reads in terms of Pauli matrices
\bel{2-15}
H^{II} = - \sum_{x\neq y} p(x,y) (s^+_x s^-_y + s^+_x s^+_y - n_x).
\ee
This can be derived in a more formal way using the infinite rate formalism
of \cite{sch}. 
The two systems I and II defined by (\ref{2-14}), (\ref{2-15}) 
are subject of this paper. The infinite rate model 
(\ref{2-15}) will be studied in some detail in one dimension with nearest 
neighbour events ($p(x,y)=0$ for $y\neq x\pm 1$). 

Since the process conserves particle number modulo 2, it splits into two
distinct subsectors corresponding to even and odd numbers of particles.
For the two state process (\ref{2-15}) the operator 
\bel{2-16}
Q=\prod_{k=1}^L
\sigma_k^z = (-1)^N
\ee
has eigenvalue $\pm 1$ in the even (odd) sector respectively. Hence
$P^\pm = (1\pm Q)/2$ projects on these sectors. Corresponding to this
separation of the dynamics it is convenient to use
\bel{2-17}
\bra{s}^{even,odd} = \bra{s} P^{\pm}
\ee
instead of $\bra{s}$ for the calculation of expectation values.

\section{Similarity transformations for DLPA}

We want to consider equivalences between model I and other systems of particles 
hopping on a lattice and which, independently of the hopping process, undergo 
some reaction when they meet on the same lattice site. I.e. we are looking for
a stochastic process of the form $\hat{H} = \hat{H}_D + \hat{H}_R$ such that 
$\hat{H} = {\cal B} H {\cal B}^{-1}$ and where $\hat{H}_D$ describes hopping on 
non-interacting 
particles and where $\hat{H}_R=\sum_x \hat{h}_x$ describes some on-site chemical
interaction. Attempting to identify all possible equivalent systems appears to
be a hopeless enterprise which is the reason why we restrict ourselves to 
equivalent processes with only on-site interaction. Therefore we assume 
${\cal B}$ to 
factorize, ${\cal B} = B_1\otimes\dots \otimes B_{\cal L}$. This has a 
consequence for the choice of $B_x$ itself: The transformed
hopping Hamiltonian $\hat{H}_D$ should not contain two-site operators other than
the hopping matrices $a^+_xa^-_y$ as such objects not describe
non-interacting particles or strictly local interactions as demanded.
This requirement is met by a matrix 
$B_x=\exp{(\alpha_x a^-_x + \beta_x a^+_x + \gamma_x n_x + \delta_x)}$ 
with arbitrary parameters $\alpha_x,\beta_x,\gamma_x,\delta_x$.
Since $\bra{s}$ has to remain invariant under the transformation, one has to
choose $\gamma_x = -\alpha_x$ and $\delta_x = -\beta_x$. This leaves a 
family of transformations with $2{\cal L}$ free
parameters.

Using the harmonic oscillator algebra one shows that 
\bea
\label{3-1}
B a^- B^{-1} & = & e^\alpha a^- + \frac{\beta}{\alpha} \left(1-e^\alpha\right)\\
\label{3-2}
B a^+ B^{-1} & = & e^{-\alpha} a^+ + 1-e^{-\alpha}
\eea
which are necessary for the calculation of $\hat{H}$. Applying the 
transformation to $H^I_R$ shows that
positivity and reality of the transformed rates constrains the 
transformation parameters to be in the domain $- \ln{2} \leq \alpha_x \leq 0$,
$\beta_x = 0$. Any transformation with $\beta_x \neq 0$ leads to events with 
negative
rate. Even though this can be compensated by inclusion of particle creation
events in the original Hamiltonian, we do not consider this possibility 
in the present work.\footnote{We just note for illustration that inclusion of 
single 
particle annihilation $A \to \emptyset$ and pair creation $\emptyset \to 2A$ 
with suitably chosen rates is equivalent to a process with pair annihilation, 
branching $A \to 2A$ and creation $\emptyset \to A$ under a transformation
with $\beta \neq 0$.} If $\beta=0$ the hopping part $H^I_D$ is invariant under
the transformation if one chooses the transformation parameters 
$\alpha_x=\alpha$ space-independent. Thus a transformation meeting all the
requirements discussed above
gives rise to the one-parameter family of processes
\bea
\hat{H}^I & = & - \sum_{x\neq y \in S} p(x,y) (a^-_x a^+_y - n_x) \nonumber \\
\label{3-3}
& &  - \sum_{x \in S} \tilde{\lambda}(x) \left[ 
(1-b + b a^+_x)(a^-_x)^2  - n_x(n_x-1) \right].
\eea
where particle either annihilate $2A \to 0$ with rate $(1-b)\lambda_x$ 
or undergo a fusion reaction $2A \to A$ with rate $b\lambda_x$
where $b=2(1-\exp(\alpha))$. This shows
that pair annihilation and fusion are in the same universality class 
irrespective
of the lattice on which the system is defined and also irrespective of the
physical environment specifying the rates $p(x,y)$ and $\lambda(x)$.
The choice of lattice is completely arbitrary.
The equivalence of the two processes was first discovered by Krebs et al. 
\cite{Kreb95} for a model with site-exclusion and 
homogeneous nearest neighbour hopping in one dimension. Universality for finite
reaction rate in any dimension was shown by Peliti \cite{Peli85} using the RG 
approach.

Since the mapping defined by the transformation ${\cal B}$ is exact it is
worthwhile to investigate some of the consequences for expectation values
for the two processes. First we note that both $\bra{s}$ and ${\cal B}$
factorize and therefore $\bra{s}B=\bra{s}$. To proceed,
we consider as initial state uncorrelated random initial 
conditions with site-dependent density $\rho_x$, i.e. we assume that at time
zero particles have been distributed randomly and independently at each site
with a Poisson distribution $P_x(n) = \exp{(-\rho_x)} \rho^n_x/ n!$. This can
be achieved by filling an originally empty lattice by the process
$H = -\sum_x \rho_x/\tau (a^+_x -1)$ up to a time $t=\tau$. Then the creation
is switched of and annihilation and fusion take place. For the pure annihilation
process the concentration $\rho_y(t)$ of particles at site $y$ is given 
by  $\rho_y(t) = \bra{s} n_y \exp{(-H^It)} \ket{\rho}$. In order to relate
the density 
\be
\hat{\rho}_y(t) = \bra{s} n_y B^{-1} e^{-H^It} B \ket{\rho}
\ee
for the mixed process to $\rho_y(t)$ we have to
study the effect of the transformation on both $\ket{\rho}$ and $n_y$.
A short calculation shows that $B \ket{P_0(x)}$ remains a Poisson 
distribution, but with density $\hat{\rho}_x = \rho_x \exp{(-\alpha)}$.
On the other hand, (\ref{2-10}) together with (\ref{3-1}) yields
$\bra{s} B n_y B^{-1} = \bra{s} a^-_y B^{-1} = \exp{(\alpha)}\bra{s} n_y$.
This gives
\bel{3-4}
\hat{\rho}_y(t) = e^\alpha \rho_y(t)
\ee
where the initial densities for the mixed process are related to the initial
density of the pure annihilation process by the relation 
$\hat{\rho}_x = \rho_x \exp{(-\alpha)}$. For a $k$-point correlation function 
$F(x_1,\dots,x_k;t) = \exval{n_{x_1}(t) \dots
n_{x_k}(t)}$  where all $x_i$ are pairwise different, (\ref{3-4}) generalizes
to 
\bel{3-4a}
\hat{F}(x_1,\dots,x_k;t)=\mbox{e}^{k\alpha}F(x_1,\dots,x_k;t).
\ee
If two or
more coordinates are equal, one has to use (\ref{3-1}) and (\ref{3-2}).
For the two-point function $F_2(y;t)=\exval{n^2_y (t)}$ this yields
\bel{3-5}
\hat{F_2}(y;t) = e^{2\alpha}F_2(y;t) - e^{\alpha}(1-e^{\alpha}) \rho_y(t).
\ee
It is interesting to consider the ratio $\Delta(t) = 
\exval{N^2(t)} - \exval{N(t)}^2/\exval{N(t)}$ of the particle number 
fluctuations to the particle number $\exval{N}=\sum \exval{n_x}$ at time $t$. 
For the
mixed process one obtains the exact relation $\hat{\Delta}(t) = \exp{(\alpha)}
(\Delta(t)-1) + 1$. This relation can  be used to measure 
the branching ratio of an experimental system, if one can measure both the
average particle number and its fluctuations and if $\Delta$ for the
pure annihilation process can be calculated analytically
for the respective physical environment.
In the simplest case of infinite pure annihilation and
homogeneous nearest neighbour hopping in one dimension this quantity can
be calculated exactly \cite{sch} and is for sufficiently large times independent 
of time and of the initial density, $\Delta(t) \to 2 - \sqrt{2}$. 

\section{Infinite reaction limit}

The transformation of the last section is independent of $\lambda(x)$.
Therefore all the results translate with little modification into the 
dynamics of the exclusion process $H^{II}$ defined in (\ref{2-15})
which describes the dynamics with infinite reaction rate.
One just has to replace $a^{-} \to s^+$ in the transformation $B$, the
results concerning the relations between correlation functions remain unchanged.

However, this simple observation is not all there is to say about this limit.
It is known that in one dimension with nearest neighbour hopping DLPA
is related to zero-temperature Glauber dynamics by a domain-wall duality
transformation \cite{Racz85}. This transformation was shown recently to
be an invertible similarity transformation \cite{Sant97}. On the other hand, 
it was noted that zero-temperature Glauber dynamics can be brought by another 
similarity transformation into a form which is the {\em transpose} of the 
Hamiltonian for DLPA \cite{Henk95}. From this we conclude that there must
be a matrix ${\cal D}$ such that 
\bel{4-1}
H^{II} =  {\cal D}^{-1}(H^{II})^T {\cal D},
\ee 
i.e. DLPA in
one dimension with nearest neighbour hopping is self-enantiodromic in addition
to being equivalent to the mixed pair-annihilation/fusion process.\footnote{
Strictly speaking, this applies only to the sector with an even number of
particles. For an odd number of particles the situation is slightly more
complicated.}

From now on we consider only the one-dimensional process on a ring with $L$
sites and with
arbitrary nearest neighbour hopping rates $l_k = p(k,k-1)$ and $r_k=p(k,k+1)$.
In order to avoid unnecessary technical complications with boundary terms, we
consider only the sector with an even number of particles.
Combining the results of Refs. \cite{Sant97,Henk95} yields
\bel{4-2}
{\cal D} = \gamma_1 \gamma_2 \dots \gamma_{2L-1}
\ee
where 
\bea
\gamma_{2k-1} & = & \frac{1}{2} \left[ (1+i) \sigma_k^z
- (1-i) \right] \\
\gamma_{2k} & = & \frac{1}{2} \left[ (1+i) \sigma_k^x\sigma_{k+1}^x
- (1-i) \right].
\eea
To prove the enantiodromy relation (\ref{4-1}) one notes that ${\cal D}$ is
unitary and transforms Pauli matrices as follows:
\bea
{\cal D}^{-1} \sigma_k^x\sigma_{k+1}^x {\cal D} & = & \left\{
\ba{ll} \sigma_k^z  & k \neq L \\
        Q\sigma_L^z & k = L
\ea \right. \\
{\cal D}^{-1} \sigma_{k+1}^z {\cal D} & = & \left\{
\ba{ll} \sigma_k^x\sigma_{k+1}^x  & k \neq L \\
        Q\sigma_L^x\sigma_1^1 & k = L
\ea \right. 
\eea
where $Q=\prod_{k=1}^L \sigma_k^z=(-1)^N$. For the even particle sector where
$Q=1$ the relation (\ref{4-1}) follows for constant rates $l_k = r_k = D$.

In order to apply enantiodromy to inhomogeneous systems we first
rewrite the Hamiltonian (\ref{2-15}) adapted to the present case. For hopping
rates as defined above the Hamiltonian reads
\bel{4-3}
H = - \sum_{k=1}^L (r_k h_k^+ + l_k h_k^-)
\ee
with the hopping-annihilation matrix 
$h_k^\pm = s^+_k s^-_{k\pm1} + s^+_k s^+_{k\pm1} - n_k$. Applying the 
transformation yields the process $\hat{H}= {\cal D}^{-1} H^T {\cal D}$ given
by
\bel{4-4}
\hat{H} = - \sum_{k=1}^L (r_k h_k^- + l_k h_{k-1}^+).
\ee
The enantiodromic process is of the same form (\ref{4-3}) as the original
process, but with dual hopping rates 
\bel{4-4a}
\hat{\ell}_k = r_k \; , \;\; \hat{r}_k = l_{k+1}.
\ee 
We shall refer to the environment
defined by the dual rates as to the dual environment. The sector with an
odd number of particles can be transformed in a similar way using
${\cal D}' = {\cal D} \sigma_L^x$. Some care needs to be taken in the treatment
of the boundary term. 

In order to make practical use of the enantiodromy we note
\bel{4-5}
\bra{s}^{even} n_{k_1}\dots n_{k_m}e^{-Ht} \ket{P_0} = 
\frac{1}{2^m} \bra{P_0} {\cal D} e^{-\hat{H}t} 
(1-\sigma_{k_1-1}^x\sigma_{k_1}^x)
\dots (1-\sigma_{k_m-1}^x\sigma_{k_m}^x)
{\cal D}^{-1} \ket{s}^{even}.
\ee
This may be written in the form 
\bel{4-6}
\bra{s}^{even} n_{k_1}\dots n_{k_m}e^{-Ht} \ket{P_0} = 
\bra{s} Q e^{-\hat{H}t} \ket{P'_0}
\ee
where $Q$ is determined by the original initial condition and the initial
state $\ket{P'_0}$ is determined by the set of sites $\{k_1,\dots,k_m\}$.
The importance of this result is seen in the transformation law
${\cal D}^{-1} \ket{s}^{even} = -i (i-1)^{L-1} \ket{0}$. Hence $\ket{P'_0}$
is a linear combination of initial states of at most $2m$ particles.
Since $\hat{H}$ does not have any particle creation terms
the time-dependence of the $m$-point correlation function with an
arbitrary many-particle initial state is completely determined by the
dynamics of the system with not more than $2m$ particles. In particular, the
density at site $k$ is given by the dynamics of the dual system with just
two particles placed initially at neighbouring sites $k-1,k$.

It remains to consider the dynamics of $Q$. Without specifying the original
initial distribution $\ket{P_0}$ and using (\ref{2-12}) we can always write 
$Q$ as a linear combination
of products with an even number of particle annihilation operators $s_{l_1}^+
\dots s_{l_2p}^+$ where $2p \leq 2m$ since $\hat{H}$ does not create particles.
Now one can follow standard procedure and perform a Jordan-Wigner transformation
\cite{Alca94,Schu95} by introducing $Q_k = \prod_{i=1}^{k} \sigma^z_i$ and
the fermionic annihilation and creation
operators
\be
\label{4-8} c^{\dagger}_k = s^{-}_k Q_{k-1} \; , \;\;
 c_k= Q_{k-1} s^{+}_k 
\ee
satisfying the anticommutation relations $\{c_k,c_l\}=\{c^{\dagger}_k,
c^{\dagger}_l\}=0$ and $\{c^{\dagger}_k,c_l\}=\delta_{k,l}$.
In terms of these operators $Q$ can be written as a linear combination
of products with an even number of fermionic particle annihilation operators 
$c_{l_1}\dots c_{l_{2p}}$ and the initial state is of the form
$c^\dagger_{k_1}\dots c^\dagger_{k_{2m}}\ket{0}$. The Hamiltonian $\hat{H}$ is 
bilinear in 
$c^{\dagger}_k$, $c_k$ and leads as shown in \cite{Schu95} to a linear 
time-evolution equation for $c_k(t) = \exp{(\hat{H}t)} c_k \exp{(-\hat{H}t)}$. 
The solution of the differential-difference equation obtained by taking the 
time-derivative of $c_k(t)$ is a single-particle problem, viz. the solution of 
the initial value problem of a single random walker in the hopping environment 
defined by $\hat{H}$, i.e.
\bel{4-10}
c_k(t) = \sum_l \hat{P}(k;t|l,0) c_l
\ee 
where $\hat{P}(k;t|l,0)=\bra{k} \exp{(-\hat{H}t)} \ket{l}$ is the conditional 
probability for the single-particle problem. This reduces the calculation
of the expectation value of $Q$ to the calculation of correlators of the form 
$\bra{s}c_{l_1}\dots c_{l_{2p}}c^\dagger_{k_1}\dots c^\dagger_{k_{2m}}\ket{0}$ 
at time $t=0$ which are given by the anticommutation relations and 
$c_l \ket{0}=0$.

The appearance of free fermions in this problem of stochastic dynamics of
classical interacting particles may seem surprising. However, 
to calculate the two-particle matrix transition probability 
$\bra{m,n}e^{-\hat{H}t} \ket{k,l}$ one can either use the free fermion
description, or, in a less technical way, remind oneselves of the meaning of an
annihilating random walk and the description of random walks in terms of a
sum over the canonical path space. In discrete space and time the transition 
probability (or conditional probability) for a single particle 
$\hat{P}(m;t|k,0)$ is the sum over all paths leading from $k$ to $m$, each 
weighted with its proper statistical weight given by the hopping rates and the
particular form of the trajectory. If two non-interacting particles, 
one starting at site $k$ and the other at site $l$, move, then the transition
probability that the particle which started at site $k<l$ reaches site $m<n$
and the particles which started at site $l$ reaches site $n$ at time $t$
is still the sum over all possible trajectories which connect $k$ with $m$
and $l$ with $n$, where each single trajectory has the same weight as in the
single particle case. Hence, for non-interacting particles, 
$\hat{P}(m,n;t|k,l;0)=\hat{P}(m;t|k;0)\hat{P}(n;t|l;0)$. This sum includes the
contribution of paths which cross each other. In an annihilating random walk
of otherwise non-interacting particles the contribution of all crossing paths
have to be subtracted. Since we are on a infinite, one-dimensional lattice
and both particles are identical this contribution is just the one given by
all paths which start at site $k$ and end at site $n$ (instead of $m$) and
which start at site $l$ and end at site $m$ (instead of $n$). Therefore
\bel{4-11}
\hat{P}(m,n;t|k,l;0)=\hat{P}(m;t|k;0)\hat{P}(n;t|l;0)-
\hat{P}(n;t|k;0)\hat{P}(m;t|l;0)
\ee
which is indeed what one obtains using the anticommutation relations in the
free fermion approach. The same subtraction scheme generalizes to higher order
conditional probabilities and is again conveniently captured in the free fermion
anti-commutation relations.

The contents of this section is the main result of
the paper. 
In the following section we consider some of its consequences.
Further relations can be obtained in a straightforward
manner for multi-time
correlation functions.\footnote{In fact, all the results obtained here for
equal-time correlations functions could have been phrased in more
conventional language using the notion of duality as often done in the
study of interacting particle systems.} The fact that $\hat{H}$ does not
contain particle creation terms ensures that also the calculation $m$-point 
correlators for different times is reduced to the solution of the $2m$-particle 
problem. The reduction of an equal-time $m$-point 
correlator to a $2m$-particle problem for the one-dimensional model with 
homogeneous hopping rates was first observed in \cite{Spou88} using a mapping
to a polymerization process. The new results presented here are the systematic
and explicit way of expressing this reduction, its generalization to
multi-time correlators and to non-translationally invariant systems and the
final reduction to a single-particle problem in the dual environment.

\section{Density decay}

The result of the previous section is completely general as far as the
hopping rates, observables and initial states are concerned. To be more
specific we consider now as initial distribution an uncorrelated random initial 
state with density 1/2 in the sector of even particle number. This distribution
is represented by the vector $\ket{1/2} = 1/2^{L-1} \ket{s}^{even}$. We study
the decay of the density in a so far unspecified, but fixed environment.
One has $\bra{1/2} {\cal D} = i (-1-i/2)^{L-1} \bra{0}$ and thus
\bel{5-1}
\rho_k(t) = \frac{1}{2} \bra{2} e^{-\hat{H}t} \ket{k-1,k}
\ee
where $\bra{2}=\sum_{n>m}\bra{m,n}$ is the sum over all states with two 
particles. Hence {\em the density at site $k$ at time $t$ for a random initial
distribution with density 1/2 is equal to one half of the survival probability 
of finding two particles anywhere in the dual environment, where initially two 
particles have been placed at sites $k-1,k$.}
The product form (\ref{4-11}) of the two-particle conditional probability 
allows us to write
\bea
\rho_k(t) & = &
\frac{1}{2} \sum_m \hat{P}_{m,k}\left[\hat{P}_{m,k}+\hat{P}_{m,k-1}\right] 
\nonumber \\ 
\label{5-4}
 & & - \sum_m \sum_{n=0}^{\infty} 
\hat{P}_{m,k}\left[\hat{P}_{m-n,k}-\hat{P}_{m-n-1,k-1}\right]
\eea
with $\hat{P}_{m,k} \equiv \hat{P}(m;t|k;0)$ as a short hand for the conditional
probability.

The expression (\ref{5-4}) contains a double sum. This can be reduced to a 
single sum by considering the time derivative for the density for which one
gets
\bel{5-3}
\frac{d}{dt} \rho_k(t) = - \frac{1}{2} \sum_m (\hat{r}_{m-1} + 
\hat{\ell}_{m}) 
\bra{m-1,m} e^{-\hat{H}t} \ket{k-1,k}.
\ee

The density at the boundary of a semi-infinite
system, i.e. an infinite system with vanishing boundary hopping rates
$r_0=\ell_1=0$ has an even simpler expression,
\bel{5-2}
\rho_1(t) = \frac{1}{2} \sum_{n=1}^\infty  \hat{P}(n;t|1,0).
\ee
In the dual system a particle which has moved to site 0 cannot escape
from there (except by annihilation with a second particle hopping from
site 1 to site 0). Thus the boundary density at time $t$ for a random initial
distribution with density 1/2 is equal to one half of the survival probability 
of a single particle anywhere in the dual environment which has an absorbing
boundary site 0 and where initially the particle has been placed at sites $1$.

Eqs. (\ref{5-1}) - (\ref{5-2}) are exact and valid for any fixed environment.
For further analysis we make two assumptions on the behaviour of the
random walker.
First consider site-symmetric processes with $r_k=\ell_{k}\equiv s_k$.
In this case, the dual process is bond symmetric, $\hat{r}_k=\hat{\ell}_{k+1}$,
and therefore $ \hat{P}_{m,k}=\hat{P}_{k,m}$. For environments such that
$\lim_{t \to \infty} \sum_n \left[\hat{P}_{m-n,k}-\hat{P}_{m-n-1,k-1}\right]/
\hat{P}_{m-n,k} \to 0$ the second term in (\ref{5-4}) becomes small
compared to the first  for large times. This expresses the density at site $k$
\bel{5-6}
\rho_k(t) = \frac{1}{2} \left( 
\hat{P}(k;2t|k;0) + \hat{P}(k;2t|k-1;0) \right).
\ee
in terms of return probabilities of a random walker.
In the presence of
an average drift with drift velocity $v$ such that $\hat{P}_{m,k}=
\hat{P}_{2vt+k,m}$ the analogue of (\ref{5-6}) reads
$\rho_k(t) = (\hat{P}(k+2vt;2t|k;0)+\hat{P}(k+2vt;2t|k-1;0))/2$. 
Here $v$ may be itself be explicitly time-dependent.
If furthermore to leading order in time 
$\hat{P}(k+2vt;2t|k;0) = \hat{P}(k+2vt;2t|k-1;0)$, then
one gets the simple result
\bel{5-7}
\rho_k(t) = \hat{P}(k+2vt;2t|k;0).
\ee
The time-dependendence of the return probability depends on the
environment.

\section{Conclusions}

The main results of this paper are:\\
(1) Diffusion-limited pair annihilation and coagulation are equivalent
to each other on any lattice and for arbritrary hopping and annihilation rates.
The time-dependent density and density correlations of the mixed process
have a simple expression in terms of the same quantities of the pure 
annihilation process (see Eq. (\ref{3-4a})).\\
(2) In one dimension with nearest neighbour hopping and infinite annihilation
rate the $m$-point correlation function for an arbitrary $N$-particle
initial state is given by correlation functions for a system with dual
hopping rates (\ref{4-4a}) with at most $2m$ particles initially. These
correlation functions can be expressed in terms of single-particle conditional
probabilities (see (\ref{4-5}), (\ref{4-10})), (\ref{4-11})). 
The derivation rests on
a free fermion formulation of path integrals for annihilating random
walks in one dimension. 
(3) Specifically for  random initial conditions with density 1/2
one gets both the exact and
approximate results of Sec. 5 the particle
density.\\
(4) For  environments satisfying the assumptions detailed in Sec. 5
the density at site $k$ at time $t$ with a random initial state is equal to the 
return probability of a single particle in the dual environment at time $2t$
(see (\ref{5-7})).\\

Expressing the density in terms
of the return probability has been done for computational convenience,
not for physical reasons. Clearly, this relation is correct only in one
dimension and only under the conditions outlined above. 
An open question which needs further investigation is the classification of 
hopping environments where these assumptions hold. It is also of interest
to study universality of the density amplitude with respect to the initial 
density. This should be relatively straightforward in the approach
developed above and is under investigation. A promising application
of our results are disordered systems \cite{Schu97,Card97} where further
analysis requires
taking a disorder average over products of single-particle conditional
probabilities. Random walks in random environments are well-studied
\cite{Alex81,Haus87} and, returning in our discussion to real systems
such as  exciton
annihilation on polymers, one expects from this knowledge to gain 
considerable insight into the behaviour of diffusion-limited
annihilation in one-dimensional disordered media.

\section*{Acknowledgments}

The author would like to thank the Department of Physics, University of Oxford,
where part of this work was done for kind hospitality and for stimulating
discussions, particularly with J. Cardy and M.J.E. Richardson.

\bibliographystyle{unsrt}

\end{document}